\documentclass{ws-pp}

\def\fig#1{Fig.~\ref{#1}}
\def\lmoon{\leftmoon\!\!}
\def\SU{{\v S}{\' U}}
\def\NA{NA}
\def\GE{GE$_6$}
\def\ME{ME}
\def\NAN{NA$_N$}
\def\MEGE{ME+GE$_6$}
\def\SUNA{{\v S}{\' U}+NA}
\def\grad{$^\circ$}

\input wasyfont

\begin{document}

\markboth{L. Brack-Bernsen and M. Brack}
         {Analyzing shell structure from Babylonian and modern times}

%
\catchline{}{}{}{}{}
%

\title{ANALYZING SHELL STRUCTURE FROM BABYLONIAN\\ AND MODERN TIMES}

\author{\footnotesize LIS BRACK-BERNSEN}

\address{Institut f\"ur Philosophie, Wissenschaftsgeschichte, 
         Universit\"at Regensburg\\D-93040 Regensburg, Germany\\
         e-mail: lis.brack-bernsen@psk.uni-regensburg.de}

\author{\footnotesize MATTHIAS BRACK}

\address{Institut f\"ur Theoretische Physik, Universit\"at Regensburg\\
         D-93040 Regensburg, Germany\\
         e-mail: matthias.brack@physik.uni-regensburg.de}

\maketitle

\begin{abstract}
We investigate ``shell structure'' from Babylonian times: periodicities 
and beats in computer-simulated lunar data corresponding to those observed 
by Babylonian scribes some 2500 years ago. We discuss the mathematical 
similarity between the Babylonians' recently reconstructed method of 
determining one of the periods of the moon with modern Fourier analysis 
and the interpretation of shell structure in finite fermion systems 
(nuclei, metal clusters, quantum dots) in terms of classical closed or
periodic orbits.

\end{abstract}

\section{Introduction}

Beats are an ubiquitous phenomenon arising from the interference of 
waves oscillating with different frequencies. In classical physics,
they occur, e.g., in water or sound waves. In microscopic physics, they
appear as quantum oscillations occurring, e.g., in the form of (super-)
shell structure in finite fermion systems such as atomic nuclei, metallic 
clusters or semiconductor quantum dots. But beating amplitudes occur 
also in coupled mechanical systems when the uncoupled subsystems have 
nearly commensurable periods. One example -- and perhaps the first
studied by mankind -- is our planetary system including the earth's 
moon. In fact, the system sun-earth-moon represents the oldest
three-body problem,\cite{gumoon} which has occupied astronomers since
more than 3000 years and until today, despite all our modern mathematical 
knowledge, is not exactly solvable. 

\begin{figure}[th]
\hspace*{0.1cm}
\psfig{file=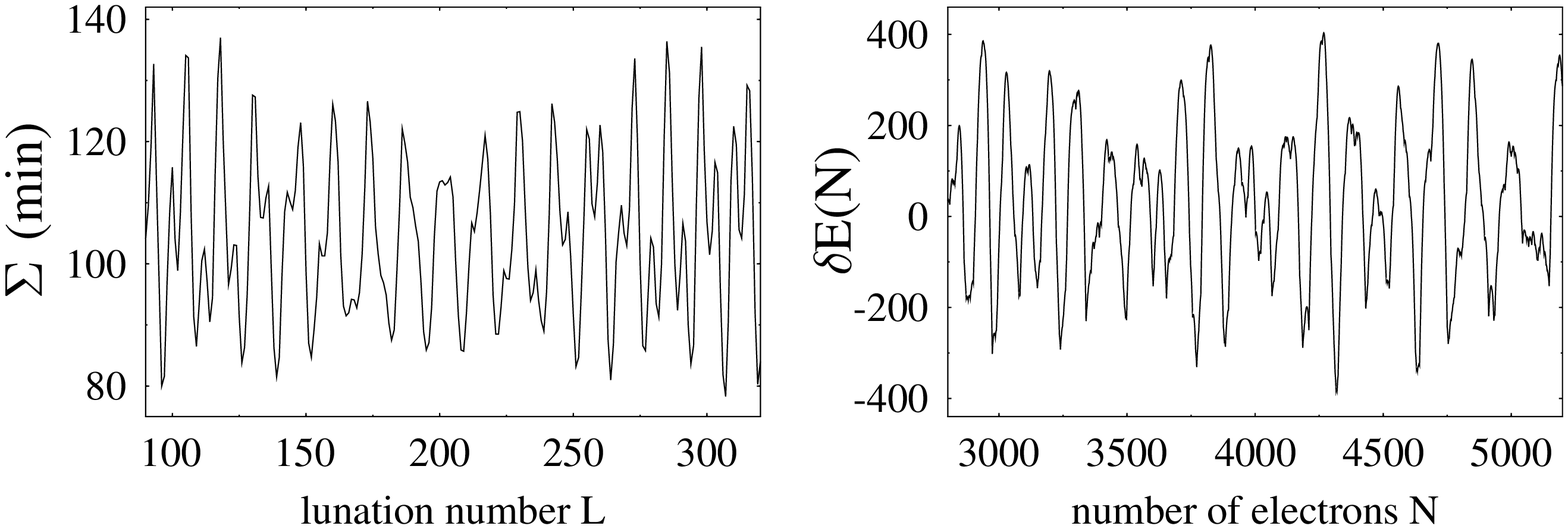,width=12.5cm}
\vspace{-0.7cm}
\caption{
{\it Left:} Sum of calculated ``Lunar Four'' $\Sigma$ = \SUNA+\MEGE, as 
observed by the Babylons since 2500 years ago, plotted versus lunation 
number $L$.
{\it Right:} Energy shell correction of the electrons in a two-dimensional 
semiconductor quantum dot, plotted versus electron number $N$.
}
\label{shells}
\end{figure}

As an illustration of the similarity of beats occurring on astronomical
and microscopic scales, we juxtapose in \fig{shells} some lunar data, 
computer simulated for the time around 500 B.C.\ in Babylon, with shell 
structure of a modern mesoscopic quantum-mechanical system. On the left 
side, we show the quantity $\Sigma$ = \SUNA+\MEGE, from which the 
Babylonians are thought\cite{lisphi} to have derived one 

\noindent
of the periods 
of the moon (as discussed in Sect.\ \ref{phi} below), calculated as a 
function of the lunation number $L$ that counts the oppositions of moon 
and sun. On the right side of \fig{shells}, we show the energy shell 
correction $\delta E$ of a two-dimensional semiconductor quantum dot 
with a radius of a few hundred nanometers,\cite{qdot} calculated as a 
function of the number $N$ of conduction electrons contained in the dot
(which was modeled here by a circular billiard).

In Sect.\ \ref{baby} we briefly discuss those lunar phenomena which were 
observed by the Babylonians and recorded on clay 
tablets. In Sect.\ \ref{modsim} we give a short account of our present
understanding, making use of modern computer simulations of the lunar
data, of some of the methods by which the Babylonians arrived at their 
precise knowledge of the periods of the sun and the moon. In Sect.\ 
\ref{four} we analyse their determination of the lunar period by applying
modern Fourier analysis. This technique is also successfully used in the 
semiclassical description of quantum oscillations in terms of classical 
closed (periodic) orbits,\cite{gutz,bablo,bbook,mbkas} which has had a 
substantial impact on recent research on quantum chaos.\cite{gubook}

\section{Observation and theory of lunar phenomena from Babylon}
\label{baby}

The Babylonian astronomy focused on special phenomena. In case of the
moon, these were eclipses and some ocurrences around conjunction (new
moon, i.e., sun and moon have the same ecliptical longitude) and 
opposition (full moon: the elongation of the moon from the sun equals 
180\grad). In case of the planets they were, e.g., the first and last 
visibility.

The so-called Babylonian mathematical astronomy, which was fully 
developed around 300 B.C., enabled the scribes to calculate positions 
and times of these special phases. They did, however, not consider them 
as functions of a continuous time variable, but they calculated a series 
of discrete phases along the path of the celestial body in question. For 
the moon they recorded times and positions of consecutive conjunctions 
or oppositions, labeled by the names of the months during which they
occurred.

\subsection{The period of the lunar velocity}

The new Babylonian month started on the evening when the crescent moon 
became visible for the first time after conjunction. The time interval 
\NAN~from sunset until moonset on that evening was observed. This 
quantity is quite obvious and easy to observe. However, from a
theoretical point of view, it is very complicated. It depends on the
velocity $v_{\lmoon}$ of the moon, its longitude $\lambda_{\lmoon}$
(position in the ecliptic), its latitude $\beta_{\lmoon}$ (distance 
from the ecliptic), and on the time interval $\Delta t$ from conjunction 
to sunset. It is an impressing achievement of the Babylonian scribes 
that they were able to develop very elegant numerical systems capable of 
calculating \NAN~for consecutive new moons, taking all the variables into 
account. The values of \NAN~were listed in so-called lunar ephemerides 
for each month. (Similarly, in the modern computer simulations discussed
in Sect.\ \ref{modsim} below, we present the data as functions of the 
lunation number $L$ counting the full moons.) At least since O. 
Neugebauer\cite{neug} we know how the ephemerides were calculated; but 
we have still rather little knowledge about how this theory was derived 
from observations.

One big question was how the period $T_{\lmoon}$ of the lunar velocity 
was found. The movement of the moon is very irregular: it keeps changing 
phase and latitude and can have its maximal velocity anywhere on the 
ecliptic. Which kind of observations did the Babylonians use? In lunar 
ephemerides concerned with conjunctions, the calculated values of the 
lunar velocity for successive new moons are given in a column called F. 
But these values are all derived from a linear zig-zag function, given 
in a column $\Phi$ appearing directly after the first column containing 
the names of the months. Therefore we must assume that $\Phi$, rather 
than F, was constructed from observations. Their common period is 
P$_{\Phi}$ = 6247/448 = 13.94420 synodic months, which is surprisingly 
accurate. P$_\Phi$ is the mean period of the lunar velocity measured on 
the days of conjunction (or at oppositions for the full-moon ephemerides). 
Let P$_{\lmoon}$ be the period of the lunar velocity, measured from day 
to day. The value of P$_{\Phi}$ corresponds to the period P$_{\lmoon}$ = 
6247/6695 synodic months = 27.55453 days. The presently known value, 
calculated for Babylonian times, is 27.55455 days.

Column $\Phi$ was long supposed to be based on lunar eclipses which 
were of great importance in Mesopotamia and had been observed since early 
times. However, the Babylonian observational records of moon eclipses were 
too inaccurate to allow for their accurate value of P$_{\Phi}$. Therefore 
it has been postulated\cite{lisphi} that some other observations were used 
to construct $\Phi$ -- namely those of some short time intervals around 
full moons.

\subsection{The Lunar Four}
\label{lunfour}

Since 747 B.C., celestial phenomena were observed regularly and recorded
month after month in the so-called Diaries.\cite{hunger} The astronomical 
observations conducted in Mesopotamia may be called the longest scientific 
project ever. Diaries were produced continuously during a period of almost 
700 years - the latest Diary found so far stems from the year 61 B.C. They 
were written in cuneiform on clay tablets: for each month of the year lunar 
phases, eclipses and planetary phases were recorded together with market 
prices, weather observations and historical events. In the earliest Diaries 
only the days of the special lunar phases were recorded.

However, starting at least back in the 6th century B.C., the Babylonians 
began to regularly observe the times between the risings and settings of 
sun and moon in the days around opposition and record the measured times. 
(The oldest preserved Diary in which \NA~is mentioned stems from the year 
568 B.C.) The following four special time intervals relating to the full 
moon, the so-called ``Lunar Four'', were observed and recorded:

\vspace*{0.1cm}
\centerline{
$\;$\SU~= time from moonset to sunrise measured at last moonset before sunrise,}
\centerline{
\NA~= time from sunrise to moonset measured at first moonset after sunrise,}
\centerline{
~\ME~= time from moonrise to sunset measured at last moonrise before sunset,}
\centerline{
\GE~= time from sunset to moonrise measured at first moonrise after sunset.}

\vspace*{0.1cm}
\noindent
These time intervals were measured in u\v{s} = time degrees (1 day = 
360\grad; 1 u\v{s} $\simeq$ 4 min), 
and since they were rather short ($<$ 20\grad~= 80 min), 
they could be measured much more accurately than the times of eclipses. 
Therefore, Lunar Four data may be much better candidates for the 
reconstruction of $\Phi$ than eclipse observations. However, these 
intervals -- all of them being similar to \NAN~-- are very complicated 
functions of the lunar velocity, its longitude and latitude, and of the 
time from opposition to sunrise: \SU~= 
\SU($v_{\lmoon}$,$\lambda_{\lmoon}$,$\beta_{\lmoon}$,$\Delta t$), etc. 
Was it possible for the Babylonians to extract information on 
$v_{\lmoon}$ from these beating functions, i.e., to find $P_\Phi$?

From cuneiform tablets it is known that the Babylonians did observe the
Lunar Four with quite some accuracy. But since only about 5\% of all
diaries have been found until now, there are large gaps in the recorded 
data. It is not possible to extract a sufficient amount of Lunar Four data 
to check exactly what information they contain. Therefore, it is of great 
help to simulate the Lunar Four data by means of a computer code for lunar 
ephemerides, as will be discussed in the following section. Since we 
thereby are concerned mainly with the partial sums \SUNA~and \MEGE~of the 
Lunar Four, it is necessary to briefly mention their astronomical 
significance.\cite{olaf,lishab} On the last morning before opposition, 
the moon sets \SU~degrees before sunrise and on the next morning, it sets 
\NA~degrees after sunrise. We see that during the day of opposition, in 
comparison to sunrise, the setting moon on the western horizon is retarded 
by the amount \SUNA. Similar arguments show that \MEGE, observed on the 
eastern horizon, is the retardation of the rising moon during the day of 
opposition.

\section{Modern simulations of old observations}
\label{modsim}

For our computer simulations of lunar observables at ancient Babylonian
times, we used a code developed by S. Moshier,\cite{mosh} which employes a 
semi-analytical lunar ephemeris adequate for historical times.\cite{chapro} 
From the risings and settings of sun and moon, evaluated at the days around
the oppositions, we computed the Lunar Four and tabulated them as functions
of the lunation number $L$. 

\subsection{The origin of column $\Phi$}
\label{phi}

\begin{figure}[th]
\begin{center}
\hspace*{-1.3cm}
\psfig{file=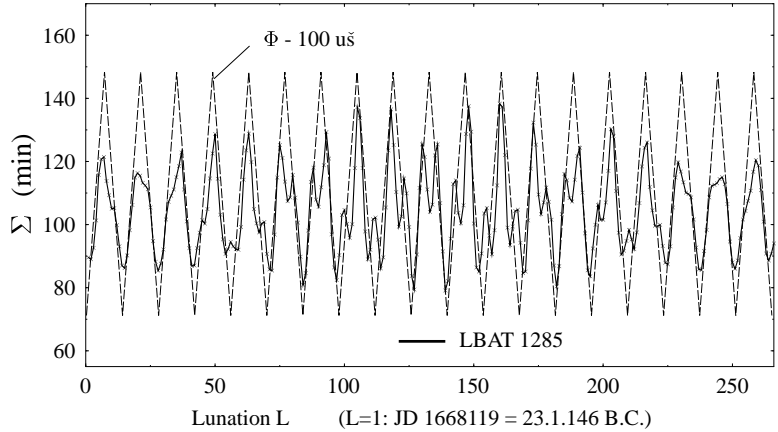,height=6.2cm}
\end{center}
\vspace{-.8cm}
\caption{
The sum $\Sigma$ = \SUNA+\MEGE~of the Lunar Four versus
lunation number $L$ (crosses and thick lines). The thin
dashed line shows the Babylonian zig-zag function $\Phi-100$\grad.
The horizontal bar covers the time span of the data on the table
LBAT 1285. $L=1$ here falls on JD 1668119 = 23.1.146 B.C.; the 
continuous counting of Julian Days starts on JD 0 = 1st of January 4713 B.C.}
\label{phifit}
\end{figure}

Using such calculated Lunar Four data and the simple idea that observations 
in the east should be combined with observations in the west, it was 
found\cite{lisphi} that the sum of all Lunar Four, $\Sigma$ = \SUNA+\MEGE, 
yields oscillations with the period $T_{\lmoon}$ that can be fitted by the 
linear zig-zag function recorded in column $\Phi$. In \fig{phifit}, 
we show the curve $\Sigma(L)$ by crosses, connected with thick lines. The
thin dashed line is the function $\Phi(L)$, shifted by an amount $-$100\grad. 
It yields an optimal fit to the calculated function $\Sigma(L)$. It
overshoots the extrema of $\Sigma(L)$ but reproduces its main oscillations
yielding, in particular the correct period P$_\Phi \simeq T_{\lmoon}$. 
Note that the phase of $\Sigma(L)$ (i.e., its position along the $L$ axis)
was not adjusted, but obtained directly from the calculated ephemerides 
appropriate for the time span (indicated by a horizontal bar in the figure)
covered by the data (shown in \fig{test1285} below) on the cuneiform tablet 
LBAT 1285. The only adjusted parameter here is the vertical shift of 
$-$100\grad, whose origin and significance have remained unclear so far.

\vspace*{-0.25cm}
In Sect.\ \ref{fourbab} we will use Fourier analysis to illustrate that
the period P$_\Phi$ can be extracted from a fit to the function 
$\Sigma(L)$, but not to any of the single Lunar Four, nor to the partial 
sums (\SUNA)$(L)$ or (\MEGE)($L$).

The conclusion is therefore nearlying that we have found the observational 
origin of the column $\Phi$ from which all data related to the lunar
velocity were derived. The hypothesis that $\Phi$ was constructed from the 
combination $\Sigma$ of lunar observables is theoretically well 
supported\cite{olaf,lishab} by the astronomical significance of $\Sigma$.

In order to support this hypothesis with historical evidence, it must be 
shown that the Babylonians really did collect Lunar Four data of
consecutive months, and that the accuracy of these data was sufficient.

\subsection{Ancient observations compared with computer-simulated data}
\label{compare}

A special type of tablets, the Goal-Year tablets,\cite{sachs} collect 
lunar and planetary data to be used for astronomical predictions in a 
special year, the ``goal year''. A Goal-Year tablet for the year $Y$ 
records all Lunar Four intervals from the year $Y\!\!-\!18$ together with 
the eclipses that took place (or were expected to occur) during the year 
$Y\!\!-\!18$. It also lists the sums \SUNA~and \MEGE~for the last six 
months of year $Y\!\!-\!19$. The Babylonians have thus, indeed, recorded 
the Lunar Four data continuously, and we can test their accuracy by 
modern simulations. 

\begin{figure}[th]
\begin{center}
\vspace*{-4.3cm}
\psfig{file=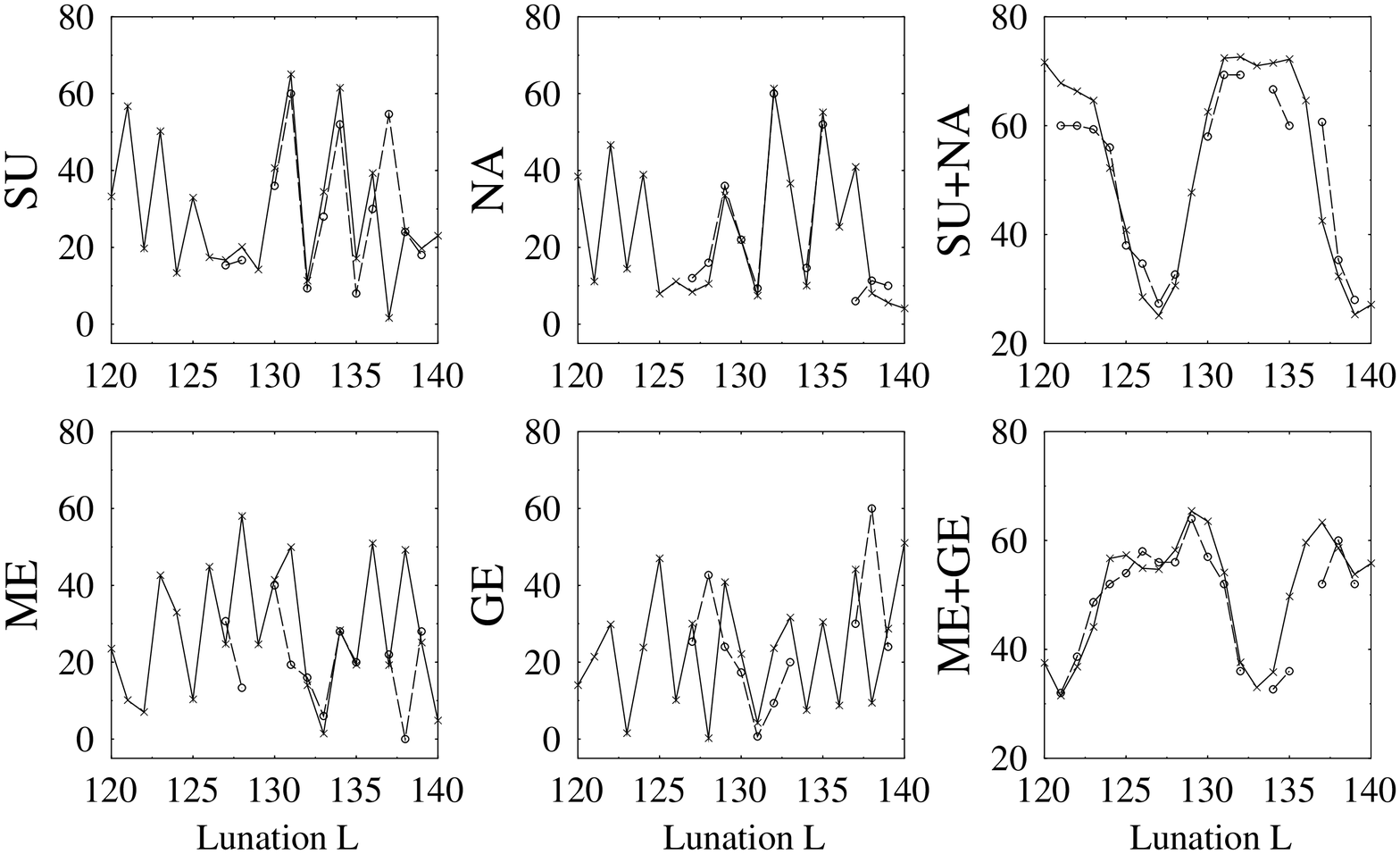,height=12cm}
\end{center}
\vspace{-1.0cm}
\caption{Comparison of Babylonian lunar data recorded on the tablet
LBAT 1285 (circles and dashed lines) with computer-simulated data 
(crosses and solid lines).$^{12}$ ($L=1$ as in \fig{phifit}.)
}
\label{test1285}
\end{figure}

In \fig{test1285} we plot the Babylonian data recorded on the Goal-Year 
tablet LBAT 1285 (circles and dashed lines) and compare them with the 
computer-simulated values (crosses and solid lines). Although the single 
Lunar Four behave rather irregularly, the agreement between old 
observations and modern calculations is very good, considering the fact 
that no adjustable parameter has been used. Especially important is the 
fact that the recorded values of the partial sums \SUNA~and \MEGE~lie 
very close to the computer-simulated curves and reflect even some of 
their fine structure. Similar agreement could be found with data recorded 
on many other tablets, one of them dating back to the times of Cambyses 
(523 B.C.).\cite{kontrol}

\vspace*{-0.4cm}
We have thus clear historical evidence that the Babylonians did record
the Lunar Four regularly over several hundred years and, in particular, 
paid attention also to their partial sums \SUNA~and \MEGE. The accuracy 
of these recorded data is sufficient to support the hypothesis that the 
lunar period P$_\Phi$ could be extracted from their sum $\Sigma$. 

\subsection{The Goal-Year method for predictions}

What if some Lunar Four data could not be observed because of bad weather
or some other reason? There must have been ways to reconstruct or predict 
them. 

\begin{figure}[th]
\begin{center}
\vspace*{-.6cm}
\hspace*{-0.9cm}
\psfig{file=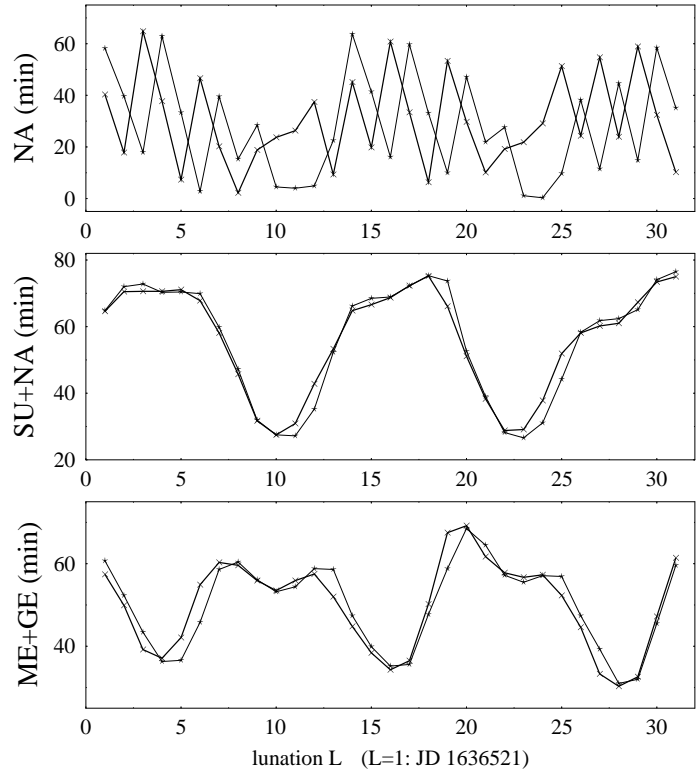,height=11.6cm}
\end{center}
\vspace{-1.2cm}
\caption{Comparison of Lunar Four data 1 Saros = 223 months apart.
Crosses and heavy lines: data evaluated at lunations $L$; stars and
thin lines: data evaluated at lunations $L\!-\!223$.
}
\label{goalyear}
\end{figure}

One Saros = 223 synodic months $\simeq$ 18 years is a well-known eclipse
cycle. A lunar eclipse observed in the year $Y\!\!-\!18$ will occur again in 
the year $Y$, and it will be visible in Babylon if it takes place during the
night when the moon is above the horizon. We can thus easily understand that 
the eclipse data on the tablet for the Goal-Year $Y$ could be used for 
predicting eclipses in the year $Y$. But the question arises: how and to what 
purpose were the recorded Lunar Four data of 1 Saros ago used? \fig{goalyear} 
helps us to answer this question. Over a period of 30 months, the functions 
\NA$(L)$, (\SUNA)$(L)$, and (\MEGE$)(L)$ are compared here with their 
respective values 223 months earlier. The sums \SUNA~and \MEGE~repeat 
themselves nicely after 1 Saros. This is not too surprising since 1 Saros, 
in a good approximation, equals an integer number of periods of $v_{\lmoon}$, 
$\lambda_{\lmoon}$ and $\beta_{\lmoon}$. However, \NA$(L)$ (taken as a 
representative of all single Lunar Four) behaves very irregularly and there 
seems to be no simple connection between \NA$(L)$ and \NA$(L\!-\!223)$. 

It is, nevertheless, possible to predict single Lunar Four data by means 
of Goal-Year data. The knowledge that (\SUNA)$(L)$ = (\SUNA)$(L\!-\!223)$ and 
(\MEGE)$(L)$ = (\MEGE)$(L\!-\!223)$ can be combined with our knowledge that 
the three variables $v_{\lmoon}$, $\lambda_{\lmoon}$, and $\beta_{\lmoon}$
will have approximately the same magnitudes at two oppositions $O_{L}$ and 
$O_{L\!-\!223}$, situated one Saros apart. The only variable determining
the Lunar Four which has changed after one Saros is the time at which
opposition takes place: 1 Saros = 223 synodic months = 6585 + 1/3 days. 
The time of opposition, compared to sunrise and sunset, is shifted by 1/3 
day. These considerations led to the following proposal\cite{mit} how the 
Goal-Year tablets could be used for predicting the Lunar Four by what we 
call the ``Goal-Year method'':
\begin{eqnarray}
\breve{S}\acute{U}(L) & = & \breve{S}\acute{U}(L\!-\!223) +
               1/3\, (\breve{S}\acute{U}+N\!A)(L\!-\!223)\,, \nonumber\\
N\!A(L) & = & N\!A(L\!-\!223) - 1/3\, (\breve{S}\acute{U}+N\!A)(L\!-\!223)\,,
               \nonumber\\ 
M\!E(L) & = & M\!E(L\!-\!223) + 1/3\, (M\!E+GE_6)(L\!-\!223)\,,\nonumber\\
GE_6(L) & = & GE_6(L\!-\!223) - 1/3\, (M\!E+GE_6)(L\!-\!223)\,.\nonumber    
\end{eqnarray}                    
The shift of 1/3 day in the time of opposition lets \SU$(L)$ become 1/3 
(\SUNA) larger than \SU$(L\!-\!223)$, while \NA~is reduced by the same
amount. The quantity \SUNA~measures the retardation of the setting moon
during the day of opposition. The correction of \SU~and \NA~by one third 
of this quantity therefore takes into account the retardation of the moon 
after 1 Saros.

The cuneiform tablet TU 11, which contains astrological as well as
astronomical sections (at the time the two were not distinguished),
nicely comfirms the Goal-Year method. In section 16 of TU 11 we find 
parts of the equations above spelled out in words.\cite{tu11} This proves
that the Babylonians had, indeed, found and used the above relations for 
the prediction of Lunar Four time intervals.

\begin{figure}[th]
\vspace*{-0.4cm}
\begin{center}
\hspace*{-0.3cm}
\psfig{file=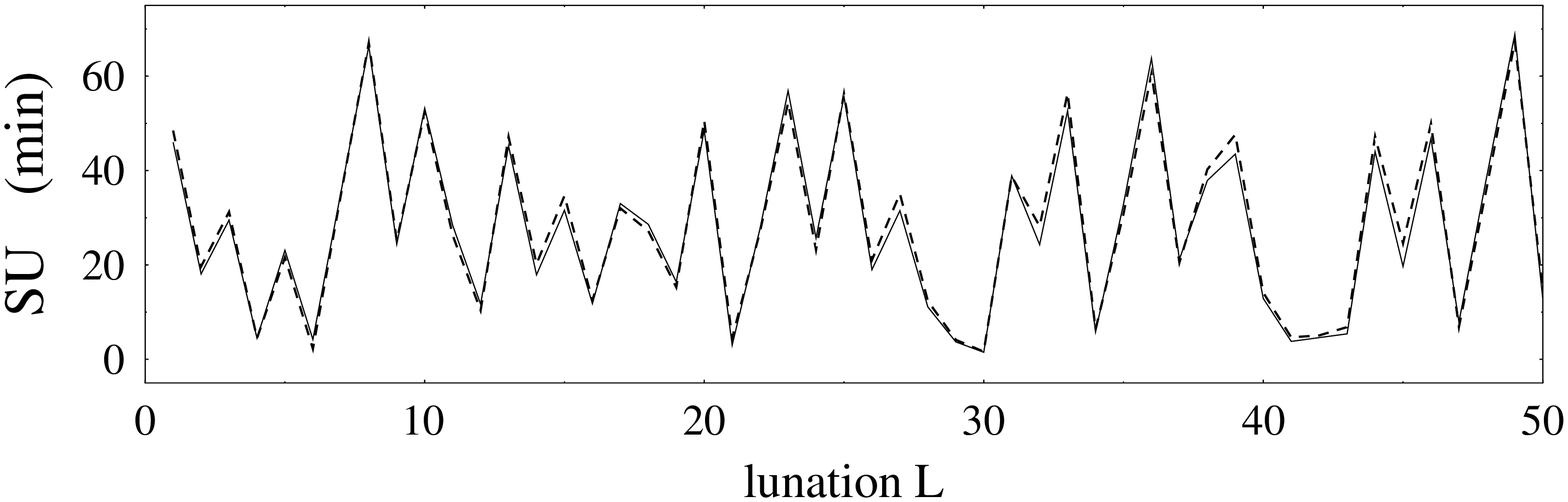,width=12.45cm}
\end{center}
\vspace{-0.5cm}
\caption{
Numerical test of the Goal-Year method for predicting \SU$(L)$ for 50
successive lunations between 236 and 232 B.C. The quantity \SU$(L)$
to be predicted is shown by the solid line; its prediction based on 
earlier data is shown by the dashed line. 
}
\label{faust}
\end{figure}

What is impressing with the Goal-Year method is that it is easy,
elegant and surprisingly precise. In \fig{faust} we illustrate the
accuracy of the method by comparing calculated values of \SU($L$) with
those predicted according to the right-hand side of the first equation 
above.

\section{Fourier analysis of shell structure}
\label{four}

To unravel the origin of beats, one can apply the technique of Fourier
analysis: by Fourier transforming the ocillating data with respect to
a suitable variable (e.g.\ energy or wave number), one obtains a spectrum 
(e.g.\ of periods or orbit lenghts) of the interfering sources. As 
mentioned in the introduction, this tool is employed also in the study of 
quantum chaos in order to establish the semiclassical correspondence of 
quantum mechanics and classical mechanics. We first give two examples of 
such analyses from atomic and cluster physics and then apply the Fourier 
analysis to the Babylonian lunar observables. (Further examples relevant 
for nuclear physics are given by Arita in this volume.\cite{arita})

\subsection{Examples of modern spectra}

\begin{figure}[th]
\begin{center}
\hspace*{-0.2cm}
\psfig{file=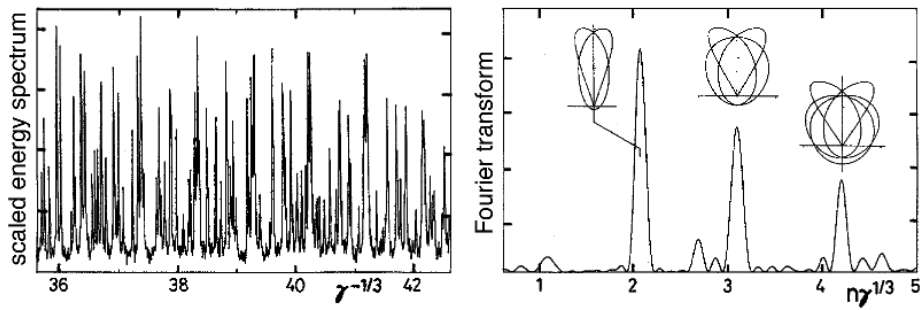,width=12.2cm}
\end{center}
\vspace{-0.5cm}
\caption{
{\it Left:} Scaled photoionization energy spectrum of hydrogen in a 
strong magnetic field. {\it Right:} Its Fourier transform. The peaks
correspond to the shown classical closed orbits of the 
electron.$^{20}$
}
\label{holle}
\end{figure}

When hydrogen atoms are put into strong magnetic fields, their 
electronic motion becomes classically chaotic. A -- by now famous --
semiclassical analysis of photoionization energy spectra of hydrogen
in magnetic fields by Holle {\it et al.},\cite{holle} using an 
extension of the periodic orbit theory of Gutzwiller,\cite{gutz}
established the one-to-one correspondence of strong Fourier peaks with 
closed classical orbits of the electron, as shown in \fig{holle}. (See
Ref.\cite{holle} for the scaling variable $\gamma$ which depends on both 
energy and magnetic field strength.)

\begin{figure}[th]
\hspace*{-0.1cm}
\psfig{file=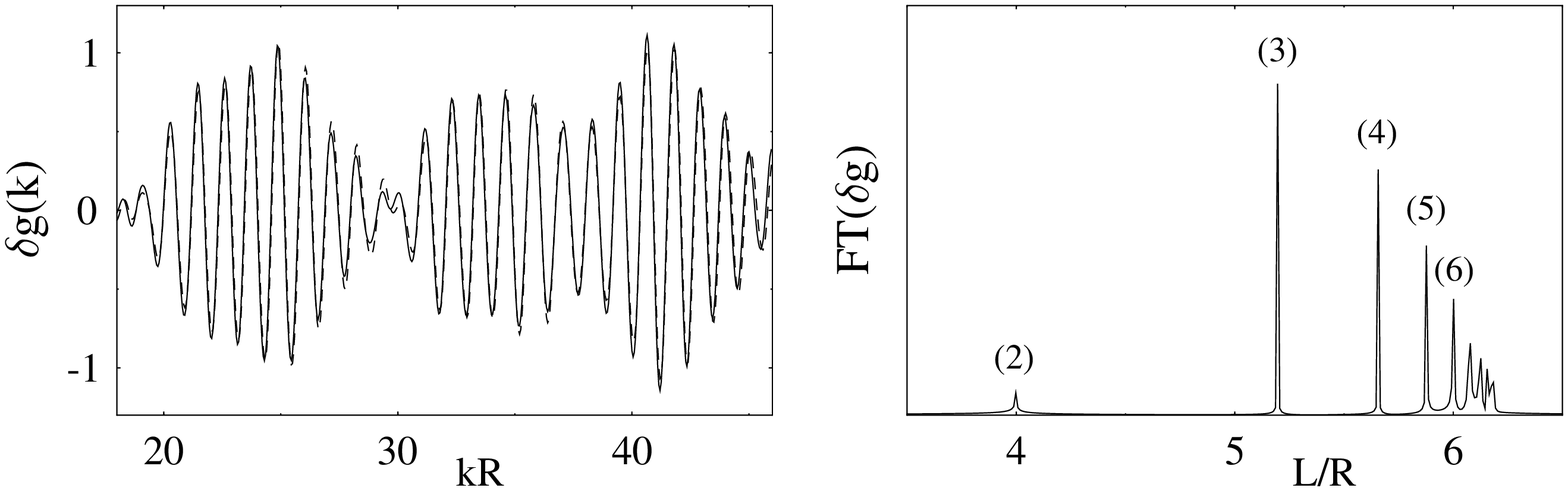,width=12.15cm}
\vspace{-0.25cm}
\caption{
{\it Left:} Coarse-grained level density $\delta g(k)$ of a spherical 
billiard (radius $R$), plotted versus wave number $kR$ (solid line: 
using the semiclassical trace formula;$^{5,6}$
dashed line: using the exact quantum spectrum).
{\it Right:} Fourier transform of $\delta g(k)$ (absolute value in 
arbitrary units) versus orbit length $L/\!R$. In parentheses are
given the numbers of corners of the periodic orbits (polygons).
}
\label{sphere}
\end{figure}

\vspace*{-0.2cm}
A textbook example\cite{bbook} of a classically integrable system
without chaos is shown in \fig{sphere}, where we display the
coarse-grained level density of a three-dimensional spherical quantum
billiard as a function of the wave number $k=\sqrt{2mE}/\hbar$, 
Gauss\-ian-averaged over a range $\Delta k=0.4/R$ in order to emphasize 
the gross-shell structure. Its Fourier transform exhibits the length 
spectrum of the shortest classical periodic orbits contributing at this 
level of resolution, whose shapes are polygons with $n$ corners (the 
number $n$ is given near the Fourier peaks in the figure). We see that 
the quantum beats in $\delta g(k)$ are mainly due to the triangles and 
squares (and, with less weight, pentagons etc.; see Ref.\cite{bbook} for
details). Although this appears to be a rather naive toy model, it is 
realistic enough to describe the supershell structure 

\noindent
that has been 
observed in the abundance spectra of metal clusters.\cite{mbclu} 

\subsection{Fourier Analysis of Babylonian observables}
\label{fourbab}

The three-body system sun-earth-moon is not integrable, but it is
-- luckily -- not chaotic. We refer to a recent review by
Gutzwiller\cite{gumoon} for an account of the various levels of
sophistication at which it has been treated over the last two millenia, 
and for the basic periods of the sun and the moon (as observed from the 
earth) which govern its observables. As we have already mentioned in 
Sect.\ \ref{baby}, the Babylonians were able to extract the period 
$T_{\lmoon}$ of the lunar velocity by a suitable combination of 
observations on the western and eastern horizon. In the following we 
shall illustrate their hypothetical procedure with the help of Fourier 
transforms.

\begin{figure}[th]
\hspace*{-0.65cm}
\psfig{file=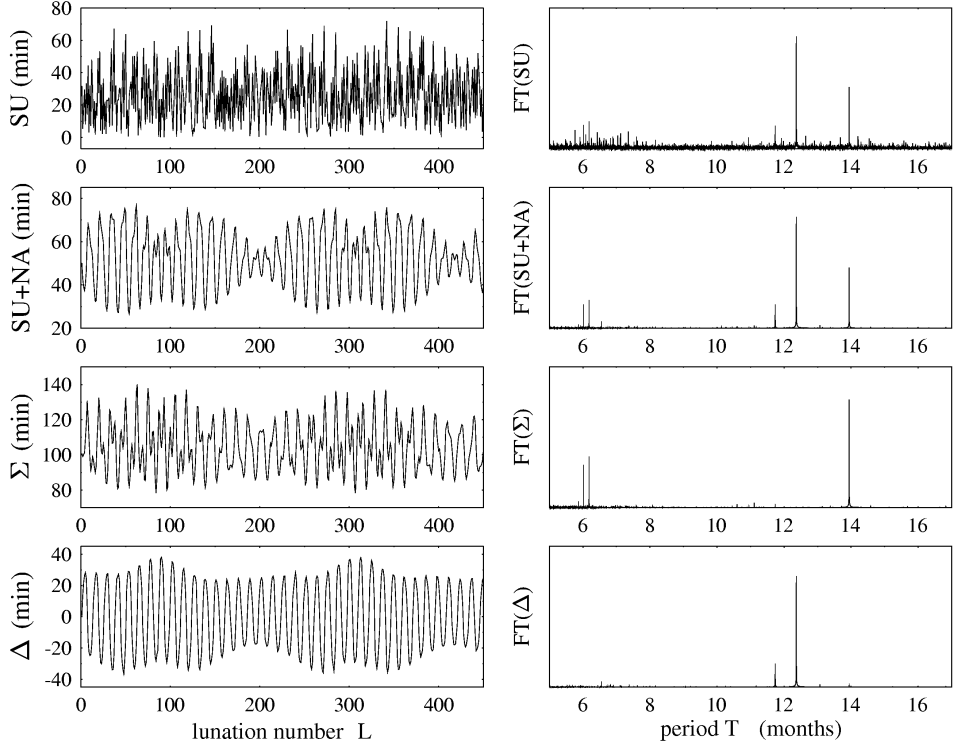,width=13.6cm}
\vspace{-0.6cm}
\caption{Calculated lunar data (from top to bottom) SU, \SUNA,
$\Sigma$ = (\SUNA)+(\MEGE) and $\Delta$ = (\SUNA)$-$(\MEGE).
{\it Left:} values as functions of the lunation number $L$.
{\it Right:} their Fourier transforms (absolute values in arbitrary
units) versus period $T$ in months.
}
\label{babfft}
\end{figure}

In \fig{babfft} we show on the upper left side two characteristic
observables recorded by the Babylonians, \SU~and \SUNA, plotted versus 
the lunation number $L$ counting the successive oppositions of sun and 
moon. On the right side we present their Fourier transforms with respect 
to $L$, which yield the spectra of periods $T$ responsible for the 
oscillations and beats in these observables. All of the ``Lunar Four'' 
\SU, \NA, \GE~and \ME~defined in Sect.\ \ref{lunfour} appear as rather 
erratic functions of $L$ yielding similar, relatively noisy Fourier 
spectra (we show here only the quantity \SU~at the top.) The spectra are 
dominated by the periods of the moon, $T_{\lmoon}=13.944$ months, and of 
the sun, $T_{\odot}=12.368$ months = 1 year, but a large number of smaller 
peaks demonstrate the complexity of the system. Next from above we show 
the sum of the two quantities observed on the western horizon, \SUNA. This 
quantity -- and similarly the sum \MEGE~observed on the eastern horizon -- 
is a much smoother and more regular function of $L$. As its Fourier 
spectrum reveals, it is mainly a beat due to the two periods $T_{\odot}$ 
and $T_{\lmoon}$. The two small components with $T\simeq 6$ months are 
responsible for the fine structure in (\SUNA)($L$) and do not affect the 
mean spacing of the ``shells'' nor the period of the beating amplitude. 
The function (\MEGE)($L$) has an almost identical Fourier spectrum, but 
its oscillations as functions of $L$ are phase shifted with respect to 
those in (\SUNA)($L$). 

This behaviour is explained if one writes (\SUNA)($L$) and (\MEGE)($L$) as 
a sum and a difference, respectively, of two periodic functions:
$$
(\breve{S}\acute{U}\!+\!N\!A)(L) = f_{\lmoon}(L) + g_{\odot}(L)\,,
\qquad            (M\!E+GE_6)(L) = f_{\lmoon}(L) - g_{\odot}(L)\,,
$$
whereby $f_{\lmoon}(L)$ does not depend on $T_{\odot}$ and $g_{\odot}(L)$
does not depend on $T_{\lmoon}$ (see Ref.\cite{olaf} for the astronomical
justification of this statement).
Hence, by constructing the sum $\Sigma$ = (\SUNA)+(\MEGE) one can eliminate
the component with the period $T_{\odot}$ of the sun and obtain a curve 
that is dominated by the lunar period $T_{\lmoon}$. Alternatively, the
difference $\Delta$ = (\SUNA)$-$(\MEGE) yields a curve dominated by the 
solar period $T_{\odot}$. (There is no evidence, however, that the Babylonians
were interested in $\Delta$.) These facts are clearly revealed by the Fourier 
spectra of
$\Sigma(L)$ and $\Delta(L)$ shown in the lower half of \fig{babfft}.
Since for both sums \SUNA~and \MEGE~the Fourier peak corresponding
to the solar period $T_{\odot}$ is much stronger than that belonging to
the lunar period $T_{\lmoon}$, both these functions oscillate mainly with
the period of the sun. It is only the function $\Sigma(L)$ that can be
fitted by the zig-zag function $\Phi(L)$ with the period of the moon, as
demonstrated in \fig{phifit}. 

\section{Summary and conclusions}

This paper has been stimulated by the close similarity between beating 
oscillations appearing in the shell structure of nuclei and other 
many-fermion systems, and in the computer-simulated lunar observables 
from Babylonian times. We have focused on the method of Fourier 
analysis for extracting the dominating periods behind beating
oscillations. For the quantum oscillations in fermionic systems this 
method is successfully employed in their semiclassical interpretation 
in terms of classical closed or periodic orbits through trace formulae, 
for which we have presented some examples.

We have given a short introduction into observation and theory of lunar
phenomena from ancient Mesopotamia. We have then shown how the use of 
modern computer simulations of the lunar and solar ephemerides at ancient 
times allowed for a partial reconstruction of the methods by which the 
Babylonians have arrived at their numerical lunar theory and their 
empirical schemes for prediciting lunar phases. There is no evidence at 
all that they had any theoretical understanding of the dynamics of the 
planetary system, nor any geometrical model for it. However, they were 
excellent numerical calculators and based their schemes on the collection 
and analysis of observational data over hundreds of years.

For a physicist it is a rather breath-taking experience to see old
``experimental'' data, recorded over 2500 years ago, reproduced by
calculations requiring the best present numerical knowledge of our 
planetary system. Of course, the computer code of Moshier makes use of 
some old astronomical data, without which the extrapolation of presently 
valid lunar ephemerides back to ancient times would not be possible. E.g., 
there are long-term variations in the earth's rotational velocity which 
result in a clock error $\Delta$T. For the time around 300 B.C., 
$\Delta$T could be determined from a solar eclipse recorded on the tablet 
LBAT 1285.\cite{steph} However, the lunar data reproduced in \fig{test1285} 
and on many other tablets have not been used as an input. Therefore the 
agreement seen in this figure, as well as in \fig{phifit}, gives us 
confidence into the accuracy of Moshier's code. At the same time it 
confirms the consistency of our present understanding of the empirical 
origins of the Babylonian lunar theory.

With the help of Fourier analysis we have illustrated the way by which 
the Babylonians could have determined the period P$_\Phi$ of the lunar 
velocity, contained in the column $\Phi$ of their lunar tables, from the 
sum $\Sigma$ of all Lunar Four. If this hypothesis is correct, then they 
have -- without knowing it -- performed a Fourier decomposition of their 
observed data. 

In any case -- the mere fact that the Babylonians some 2400 years ago were 
able to determine the length of the synodic month with an accuracy of six 
digits must be considered as one of the greatest scientific achievements 
of history.

\vspace*{0.3cm}

Lis Brack-Bernsen is grateful to the Deutsche Forschungsgemeinschaft for
financial support.

\end{document}